\documentstyle[11pt] {article}

\title{Covariant Formulation of the Dynamics in a Dissipative Dielectric Obtained from a
Simplified Lagrangian}
\author{A. Yahalom$^a$, R. Englman$^{a,b}$ and Y. Pinhasi$^a$\\
$^a$ College of Judea and Samaria, Ariel 44284, Israel\\
$^b$ Department of Physics and Applied Mathematics,\\
Soreq NRC,Yavne 81800,Israel\\
e-mail: asya@yosh.ac.il; englman@vms.huji.ac.il; yosip@yosh.ac.il}

\begin{document}
\maketitle

\newcommand{\beq} {\begin{equation}}
\newcommand{\enq} {\end{equation}}
\newcommand{\ber} {\begin {eqnarray}}
\newcommand{\enr} {\end {eqnarray}}
\newcommand{\eq} {equation}
\newcommand{\eqs} {equations }
\newcommand{\mn}  {{\mu \nu}}
\newcommand{\sn}  {{\sigma \nu}}
\newcommand{\rhm}  {{\rho \mu}}
\newcommand {\SE} {Schr\"{o}dinger equation}
\newcommand{\sr}  {{\sigma \rho}}
\newcommand{\bh}  {{\bar h}}
\newcommand {\er}[1] {equation (\ref{#1})}
\newcommand {\ers}[1] {equation (\ref{#1}) }
\newcommand{\gb} {{\bf \gamma}}
\newcommand{\gcrb}  {{\bf \gamma^+}}
\newcommand{\gd} {{\dot \gamma}}
\newcommand{\gcr} {{\gamma^+}}
\newcommand{\gcrd} {{ \dot \gamma^+}}
\newcommand{\ro} {{ \gamma  \gamma^+}}

\begin {abstract}
Equations of motions and energy-momentum density tensors are
obtained for a  dispersive and  dissipative medium sustaining
electric and magnetic polarizations, using Lagrangian formalisms.
A previous work on the subject by the authors has been simplified
by reduction in the Lagrangian of the number of independent vector
fields, with which the sink modes (associated with the
dissipation) interact. The paper also formulates the dynamics of
the electromagnetic field in the medium in a covariant
(relativistic) manner. We discuss (and compare) the results,
especially from the point of view of the positivity of the energy
density.

\end {abstract}


\section{Introduction}
Lagrangian methods provide a systematic and unified approach to
dynamic phenomena. Every mature subject in physics  can and should
be described by a Lagrangian. To name a few: particle physics is
described by the "standard Model" Lagrangian, General relativity
by the "Curvature Scalar" Lagrangian and so on.

Nevertheless, there are some phenomenological
treatments that are not obviously expressible by a Lagrangian
formalism and have led to ambiguities and controversies. An
instance of this is the dynamics of polarization in
dispersive-dissipative materials. While the equations of motion
(shown below in \er{eleom} and \er {mageom}) have been in frequent
use, the derivation of these equations from a Lagrangian has been
wanting for some time. As a consequence, basic quantities like
energy and momentum densities, that are integral parts of the
Lagrangian method, have become variously formulated and subject to
dispute. A brief account of the history of energy densities is
found in \cite{EnglmanY}. Following that account and adopting a
Lagrangian formulation, that work has unambiguously derived an
expression for the energy density.

Relativistic treatments of dissipative electromagnetic media, but
not based on a Lagrangian formalism, were given in
\cite{KluitenbergG}. A comparison of three different approaches
for relativistic treatments of dissipative electromagnetic media
was undertaken in a book by Fano, Chu and Adler \cite{FCA}. In
this book the authors compared the approach they endorse with two
other approaches, denoted the Minkowski and Amperian approaches.
The Minkowskian approach was criticized by the authors because it
assumes a linear scalar relation between the displacement field
$\vec D$ and the electric field $\vec E$, and between the magnetic
field $\vec H$ and the magnetic flux density field $\vec B$. The
authors argue that this relation is not invariant under a general
Lorentz transformation and thus is not objective but subjective
and can only be regarded as true in a a particular frame of
reference. Furthermore, even according to an observer in rest with
respect the electromagnetic medium, the linear scalar relations
are not generally true since some magnetic materials have a
hysteresis curve and thus $\vec B$ and $\vec H$ are not linearly
related, while other materials possess a tensorial dielectric
constant instead of a scalar one. Fano, Chu \& Adler also claim
that the Lorentz force resulting from the Minkowskian formalism
contradicts experimental evidence regarding polarized and
magnetized materials. The Amperian approach is criticized in the
book on the basis that the resulting Lorentz force contradicts
experimental evidence regarding magnetized materials. The authors
of this paper do not follow the approach of Fano, Chu \& Adler,
since it is not based on a variational approach and thus one
cannot canonically derive an energy expression. Furthermore, Fano,
Chu \& Adler do not regard the magnetization and polarization of a
material as degrees of freedom subject to their own dynamics;
rather, these are treated as given source terms for the the
electric and magnetic fields. It should be noted however, that
both the energy density obtained by them and that using the
Amperian approach are positive.

Other approaches, notably for physical systems containing both
viscous flows and electromagnetic fields are discussed by
Felderhof \cite{Felderhof} and Rosensweig \cite{Rosensweig}. The
first work postulates an energy-momentum tensor (equation (3.12)
in \cite{Felderhof}); in our formalism we will aim at deriving the
energy momentum tensor from a Lagrangian using the well known
canonical formalism (see below). The second work \cite{Rosensweig}
is based on an "engineering" type of approach. Rosensweig has
postulated balance equations in a moving frame for linear
momentum, angular momentum and energy and has derived
"constitutive equations" using the first and second laws of
thermodynamics.  The energy momentum tensor was not considered in
this work, nor a Lagrangian to derive the energy momentum tensor
from.

It is evident that the energy density, written as $T^0_0$, is  a
component of the energy momentum tensor $T^k_j$ (\cite
{LandauLP}-\cite {Jackson}). This is uniquely derivable from the
Lagrangian density $ {\cal L}(u_k,u_{k,j})$, which is composed of
the field variables $u_k$ and of their derivatives $ u_{k,j}$. (A
symbol after the comma represents a derivative.)
 The formal definition of the energy-momentum density is given by
\beq T^k_j = \frac{\partial {\cal L}}{\partial u_{i,k}}u_{i,j}-
{\cal L}\delta ^k_j \label {Tkj} \enq Summation is implied for
repeated
  indexes. In our convention the
indexes take the value
 $0$ for the time component and the numbers $(1,2,3)$ for the remaining,
 space components.  The dot over a quantity will
 signify a time derivative.

The expression  for the energy found in \cite{EnglmanY} was
positive. The essential idea there was the introduction of added
degrees of freedom (named sinks),  which are depositories of the
the energy residing in the electromagnetic field and in the
polarization of the medium. Physically, the sinks are macroscopic
modes of motion in the solid (and less obviously, in its
surrounding) different from the polarization modes, but weakly
interacting with them. These degrees of freedom have been given
microscopic interpretation in the thermodynamic limit, in terms of
a set of interacting oscillators with a special interaction matrix
\cite{Ford}. The need to include additional degrees of freedom to
treat the irreversible behavior of polarizable systems was noted
in a classic text \cite{GrootM}. The proof that any set of differential
equations can be embedded in a larger set that can be derived from a
variational principle was supplied by Bateman \cite{Bateman}.

 Our aim in the present work is twofold: first we simplify our previous formulation by
 reducing the
  number of vector degrees of freedom. Thus, the formalism presented here
  contains two vector degrees of freedom
  less than the previous one. The resulting Lagrangian is considerably simpler without
   compromising the
  power of our formalism. Secondly, in later sections,  we give a relativistic
  covariant formalism, making the theory valid also in a moving system of reference.
  The motivation for this is twofold:
  \begin{enumerate}
    \item It is well known that any physical theory should obey the laws of special
relativity and should be invariant under Lorentz transformation -
in this paper we  construct a theory along the same lines as our
previous theory \cite{EnglmanY}, which satisfied these
requirements.
    \item Our previous theory did not contain spatial derivatives of certain quantities; the
principle of covariance has helped us put those derivatives in.
  \end{enumerate}
Further, we discuss the implications.  Some of the equations
in the earlier sections have already appeared in \cite {EnglmanY};
however, they are indispensable both for a self-contained reading
of the simplified theory and as an introduction to and comparison
with the covariant theory.

\section { Time independent polarizations }

Our first task is  to formulate the Lagrangian density of the
electromagnetic and polarization and magnetization fields
 for a material whose polarization are
static. We use CGS Gaussian units.  Connections between the
electric displacement field $\vec D$ and the electric field $\vec
E$ and polarization $\vec P$, and, likewise, between the magnetic
field $\vec H$ and the magnetic induction field $\vec B$ and
magnetization $\vec M$ are then provided by the relations: \beq
\vec D = \vec E + 4 \pi \vec P. \label{DE} \enq  \beq \vec H =
\vec B - 4 \pi \vec M. \label{HB} \enq The above fields satisfy
both the homogeneous Maxwell's equations: \beq curl \vec E +
\frac{1}{c}{\dot{\vec B}}=0, \qquad ~div\vec B=0 \label{Maxhom}
\enq and the inhomogeneous equations: \beq curl \vec H -
\frac{1}{c}\dot{ \vec D}=\frac{4 \pi}{c}\vec J, \qquad ~div \vec
D= 4 \pi \rho \label{Maxinhom} \enq We further make the assumption
that both free charges $\rho$ and currents $\vec J$ are zero.

It is not possible to derive these equations from a Lagrangian
density expressed in terms of the  fields. However, this
difficulty is traditionally overcome by representing the fields in
terms of vector $\vec A$ and scalar $\Phi$ potentials, as follows:
\beq \vec E = -\vec \nabla \Phi - \frac{1}{c} \dot{ \vec A},
\qquad \vec B = curl \vec A \label{pot} \enq Using these
definitions one can see that the homogenous \eqs (\ref{Maxhom})
 are satisfied automatically. The inhomogeneous \eqs (\ref{Maxinhom})
 are the outcome of the functional derivation of the Lagrangian below:
\ber
{\cal L} &=& {\cal L}_{EB} + {\cal L}_{PE} + {\cal L}_{MH}
\nonumber
\\
{\cal L}_{EB} &=& \frac{1}{8 \pi}[\vec E^2 - \vec B^2]
\nonumber
\\
{\cal L}_{PE} &=&  \vec P \cdot \vec E
\nonumber
\\
{\cal L}_{MH} &=&  \vec M \cdot \vec H \label{staticlag} \enr In
forming the derivatives  it is understood that all field
quantities as given in terms of the potentials of \er{pot}.  The
energy density is then obtained from  \er{Tkj}: \beq
 T^0_0 = \frac{\partial {\cal L}}{\partial \dot{\vec A}}\cdot \dot{\vec A}-
  {\cal L}.
  \label{edk1}
  \enq
After some algebra this yields for the Lagrangian of
\er{staticlag}  the result: \beq
 T^0_0 = \frac{1}{8 \pi}[\vec E^2 +  \vec H^2 +  (4 \pi \vec M )^2]
  \label{edstatic}
 \enq
The last term $ \vec M^2$ can be usually discarded for static
magnetization, since it represents nothing but a time-independent
constant. However, we include it for reference in future sections,
where the electric polarization and the magnetization will become
dynamic degrees of freedom.

  \section {Equations of motion with dissipative terms}

The equations that  treat the development of the polarization
${\vec P}$ by
   an applied electric field ${\vec E}$, and the magnetization ${\vec M}$ as
   induced by an external magnetic field ${\vec H}$ are, as appearing in  e.g.,
    \cite {Ruppin}, the following:
  \ber
  \ddot{\vec P} +\Gamma _e \dot {\vec P} + \omega^2_r {\vec P} =
  \omega^2_p {\vec E}
  \label {eleom}\\
  \ddot{\vec M} +\Gamma _h \dot{\vec M} + \omega^2_0 {\vec M} =
  F \omega^2_0 {\vec H}
  \label {mageom}
  \enr
 with all symbols and notation as defined in \cite {Ruppin}.
 These equations hold for $t \geq 0$. Fourier transformed to the frequency ($\omega$) domain,
  the equations for the frequency components of the fields are written, after
  a transposition of terms   as
  \beq
  [-\omega^2-i \omega\Gamma_e+\omega_r^2] \vec P(\omega)=\omega_p^2
  \vec E(\omega)+{\dot {\vec P}}(t=0)-(i\omega-\Gamma_e) \vec P(t=0)
  \label{transf1}
  \enq
  and similarly for the magnetization. (Here time-derivatives of the polarization were
   integrated  by parts in the manner: $\int_0^\infty{\dot P}(t)e^{i\omega t}dt=[P(t)e^{i\omega
  t}]^\infty_0-i \omega\int^\infty_0 P(t) e^{i \omega t}dt$ and neglecting  the field at
  infinity, etc. As discussed in \cite {GrootM}, there can be a polarization at $t=0$ due
   to permanent dipoles.)
  One can introduce the electric and magnetic susceptibilities, defined by
  \ber
  \chi_e(\omega) & = &
  \frac{\omega_p^2/\omega_r^2}{1-(\frac{\omega}{\omega_r})^2-i \frac{\omega\Gamma_{e}}{\omega_r^2}}
  \label{chie}\\
  \chi_m(\omega) & = &
  \frac{F}{1-(\frac{\omega}{\omega_0})^2-i\frac{\omega\Gamma_{h}}{\omega_0^2}}
  \label{chim}
  \enr
  to write the equations of motion for the frequency components as:
  \ber
  \vec P(\omega) & = & \chi_e(\omega)[\vec E(\omega)+
  \frac{{\dot {\vec P}}(0)-(i\omega -\Gamma_e)\vec P(0)}{\omega^2_p}]
  \nonumber\\
  \vec M(\omega) & = & \chi_m(\omega)[ \vec H(\omega)+\frac{{\dot
  {\vec M}}(0)-(i\omega-\Gamma_h) \vec M(0)}{F\omega^2_0}]
  \label{PM1}
  \enr
The electric and magnetic susceptibilities are complex functions with related imaginary ($\Im$) and real
($\Re$) parts.
When the electric losses are sufficiently low $\Gamma _e<<1$,
the electric susceptibility can be approximated in the vicinity of the resonance by:
\beq
\Im [\chi_e] = \frac{\omega_p^2}{\omega_r \Gamma_e} \frac{1}{(\frac{\omega-\omega_r}{\frac{\Gamma_e}{2}})^2+1},
\qquad \Re [\chi_e] = - \frac{\omega-\omega_r}{\frac{\Gamma_e}{2}} \Im [\chi_e]
\enq
Analogous relations can be derived for the real and imaginary parts of the
magnetic susceptibility for $\Gamma_h<<1$:
\beq
\Im [\chi_m] = \frac{F \omega_0}{\Gamma_h} \frac{1}{(\frac{\omega-\omega_0}{\frac{\Gamma_h}{2}})^2+1},
\qquad \Re [\chi_m] = - \frac{\omega-\omega_0}{\frac{\Gamma_h}{2}} \Im [\chi_e]
\enq
Returning now to \er{eleom} and \er {mageom}, for self-consistency
 these have to be supplemented by the Maxwell-equations given in
the previous section and
 derived from the Lagrangian of \er{staticlag}
that contain the contravariant four-vectors $A^\alpha= (\vec
A,\Phi)$.

To take account of the dissipative nature of the processes we
introduce a set of new fields, designated canonical fields. These
will be formally distinguished from
 the previous, physical fields by writing them in lower case
 symbols. The introduction of these fields follows theories by
 \cite{Hannay} -\cite{Bekenstein1982}.
    The canonical fields (taken to be real) are  made to be part of  a
  Lagrangian formulation. We note that for  the
preceding physical quantities (written in upper case symbols),
this cannot be achieved \cite{Hannay}.

To make our new notation explicit, we shall work with
 electric polarization $p^{\alpha}$ and magnetization $ m^{\alpha}$
 and
 the electromagnetic fields
$e^{\alpha}$, $h^{\alpha}$ and the (contravariant)
vector-potentials $a^{\alpha}$. It will be later demonstrated that
these new, canonical fields contain a time dependent factor which
(partly, but not fully) neutralizes the dissipative process.
Otherwise, the choice of the fields (and of the various constant
factors) is guided by the
 requirement that we recapture the form of the equations of motion and the energy
  density currently widely
 employed in solid state optics in the appropriate limiting case of no dissipation.
 (This requirement impairs at times  the formal symmetry between
 the modes linked with the polarization and the magnetization,
 respectively.)  The physical meaning of the canonical fields will be made
 clear later by comparing
their equations of motion with those of the physical fields.
(For a similar procedure see \cite{Hannay}.)

In addition, we shall introduce
 two new scalar fields, the so called "sink" fields: the first being $\Psi$, that
 incorporates macroscopic degrees of freedom associated with the dissipation of the
 polarization (analogous to the degree of freedom denoted with the same
  symbol in \cite {Bekenstein1982}) and the  second, the sink-field
  $\Phi$ which accounts for the dissipation in the  magnetic mode (and originates
   physically in atomic spin or angular momenta in the solid).

  \section {The Lagrangian with sink terms}
  \label{dissipativeLagrangian}

  The addition of sink fields requires the extension of the Lagrangian as follows:
  \beq
  {\cal L} = {\cal L}_{eb}+ {\cal L}_{pe} +  {\cal L}_{mh} +
  {\cal L}_{p\Psi} +  {\cal L}_{m\Phi} +
  {\cal L}_{\Psi} + {\cal L}_{\Phi}
    \label {Lagrangian}
  \enq
  In this Lagrangian we have first the Lagrangian of \er{staticlag},
there written in terms of the electromagnetic fields, and now
rewritten in the new lower-case variables as
  ${\cal L}_{eb}+ {\cal L}_{pe} + {\cal L}_{mh}$.
  The above expression also contains the Lagrangians in the polarization modes,
  the magnetization modes, the electromagnetic fields, the $\Psi$ and
  $\Phi$ sink fields and then the interaction-Lagrangians between
  polarization and electric field, magnetization and magnetic
  field and, finally, the polarization and magnetization with their
  respective sink fields.
  The various terms are written out as:
  \ber
  {\cal L}_{eb} &=&  \frac{1}{8 \pi}[e_1^2 - b_1^2]\\
  {\cal L}_{pe} &=&  p e_1\\\
  {\cal L}_{mh} &=& m h_1\\
  {\cal L}_{p\Psi} &=&\frac{1}{2\omega^2_p}(\dot{p}^2 -(\omega^2_r - \dot \Psi^2) p^2
   - 2 p \dot{p} \dot{\Psi})
    \label {lagrfirst}\\
  {\cal L}_{m\Phi} &=&\frac{1}{2F\omega^2_0}(\dot{m}^2 -(\omega^2_0 - \dot \Phi^2) m^2
   - 2 m \dot{m} \dot{\Phi})     \\
    {\cal L}_{\Psi} &=& {1\over{2}}\kappa{\dot{\Psi}}^2 - {1\over{2}}\rho\Psi^2
   \label{LagrPsi}\\
  {\cal L}_{\Phi} &=& {1\over{2}}\lambda{\dot{\Phi}}^2 - {1\over{2}}\nu\Phi^2
\label {LagrPhi}\\\nonumber
   \enr
 having chosen the
 $x$ (or 1) axis as the direction of the polarization and magnetization vectors .
 In the above, we have expressed  $ {\cal L}_{\Psi}$ and ${\cal L}_{\Phi}$ as
 harmonic modes of motion (when there is no coupling to other
 fields.) This choice seems to allow the simplest type of representation of unspecified (and, so
 far, arbitrary) degrees of freedom. The symbols $\kappa,\rho,\lambda,\nu$
 are constant, non-negative coefficients, whose values
 depend on the nature of the relaxation mechanisms.

 We next derive  $\cal L$ with respect to each degree of
  freedom. In accordance with the Euler-Lagrange equations, this procedure leads to the
  equations of motion.
  In detail, for the variables  $p$ and $m$ the equations of motion take the following form:
  \ber
    \ddot{p}-( \ddot{\Psi}+{\dot{\Psi}}^2 -\omega^2_r)p & = & \omega^2_p
    e_1
    \label{peom}\\
    \ddot{m}-( \ddot{\Phi}+{\dot{\Phi}}^2 -\omega^2_0)m & = &F \omega^2_0
    h_1
    \label{meom}
    \enr
The electromagnetic equations of motion are identical to Maxwell's
equations in the new (small letter) variables. Lastly, for the
sink coordinates one finds \ber \ddot
{\Psi}+\frac{\rho}{\kappa}\Psi = {1\over{\kappa
\omega^2_p}}\frac{\partial(p(\dot{p}-p\dot{\Psi}))}{\partial t}
\label {Psieom}\\
\ddot {\Phi}+\frac{\nu}{\lambda}\Phi =
\frac{1}{\lambda F \omega^2_0}\frac{\partial(m(\dot{m}-m\dot{\Phi}))}{\partial t}
\label {Phieom}
\enr

  The expression for the energy density $T^0_0$ is found from
  \er{Tkj},  namely,
\beq
 T^0_0 =  \sum_{k}  \frac{\partial {\cal L}}{\partial \dot{u_{k}}}\dot{u_{k}}-
  {\cal L}
  \label{edk}
  \enq
with the summation over all degrees of freedom. We separate  the
electromagnetic, the polarization and the magnetization  parts
\beq T^0_0=(T^0_0)_{e,h}+(T^0_0)_{p,\Psi} +(T^0_0)_{m,\Phi} \label
{separate} \enq where the first term has the well known form of
the electromagnetic energy density, already  given in
\er{edstatic}: \beq (T^0_0)_{e,h}= \frac{1}{8 \pi}[ e^2 + h^2 + (4
\pi m)^2] \label{emden} \enq The others take the forms: \ber
(T^0_0)_{p,\Psi} & = & {1\over{2 \omega^2_p}}\big[({\dot {p}-
p{\dot{\Psi}}})^2 +\omega^2_r p^2\big]  +
\frac{\kappa}{2}{\dot{\Psi}}^2+
    \frac{\rho}{2}\Psi^2\\
    & = &{1\over{2 \omega^2_p}}({\dot {p}}^2
+\omega^2_r p^2)  +    \frac{\kappa}{2}{\dot{\Psi}}^2+
    \frac{\rho}{2}\Psi^2 + {1\over{2 \omega^2_p}}p^2{\dot{\Psi}}^2\
    \nonumber\\
    & - & {1\over{2\omega^2_p}}\frac{\partial {p^2}}{\partial t}{\dot{\Psi}}
     \label {edp2}
 \enr
 and
\ber
(T^0_0)_{m,\Phi} & = & \frac{1}{2F\omega^2_0}\big[(\dot
{m}-m{\dot{\Phi}})^2+{\omega^2_0}m^2\big]
        + \frac{\lambda}{2}{\dot{\Phi}}^2 + \frac{\nu}{2}{\Phi}^2\\
& = &\frac{1}{2F \omega^2_0}({\dot {m}}^2
+\omega^2_0 m^2)  +    \frac{\lambda}{2}{\dot{\Phi}}^2+
    \frac{\nu}{2}\Phi^2 + \frac{1}{2F\omega^2_0}m^2{\dot{\Phi}}^2\
    \nonumber\\
    & - & \frac{1}{2F\omega^2_0}\frac{\partial {m^2}}
    {\partial t}{\dot{\Phi}}
\label{edm2} \enr The first-written forms are sums of squares with
non-negative coefficients; this ensures  that
 each part of the energy density is positive (non-negative). The usual
  expressions for the energy density, e.g. in \cite{Jackson} or
\cite {Ruppin}, differ from the above by the presence of the terms
in $\Phi$ and $\Psi$ and their time derivatives. In the following
section we shall eliminate these variables by making use of the
equations of motion, \er{Psieom} and \er{Phieom}.

\section{A simple solution}

To work out a fully solvable case, we postulate that the
coefficients $\rho$ and $\nu$ in  \er {LagrPsi} and \er {LagrPhi}
vanish.  Then the equations of motion for the sink-modes,
 ~\er{Psieom} and \er {Phieom},  can be integrated. We shall carry
through the electric polarization case, but an analogous
development holds for the magnetization sink-variable. We assume
the following initial
 conditions for $\Psi(t)$
 \beq
 \Psi(0)=0, ~~~~\dot{\Psi}(0)=\Gamma_e /2
 \label{ic}
 \enq
 With these choices the differential  \er{eleom} is regained for short times,
 $t<<\frac{2}{\Gamma_e}$, as will be
 shown presently. (This is similar to the procedure in \cite {Hannay}). Then,
  from \er {Psieom},
 \beq
 \ddot{\Psi}={1\over{2\kappa  \omega^2_p}}\frac{{\partial}^2}{ \partial
 t^2}{p^2(t)}-{1\over{\kappa  \omega^2_p}}\frac{\partial ({p^2(t)}\dot{\Psi})}{\partial t}
\label{Psiddot} \enq A first integration gives \beq
\dot{\Psi}={1\over{2\kappa \omega^2_p}}\frac{\partial{p^2(t)}}{
\partial
 t}- \frac{{p^2(t)}\dot{\Psi}}{\kappa  \omega^2_p}+C
 \label{Psidot}
 \enq
 which leads to
 \beq
 (1+\frac{{p^2}(t)}{\kappa  \omega^2_p}){\dot{\Psi}}={1\over{2\kappa
  \omega^2_p}}\frac{\partial {p^2(t)}}{ \partial t} + C
 \label{Psidot2}
 \enq
where the constant is \beq C=\frac{\Gamma_e}{2}
(1+\frac{{p^2(0)}}{\kappa  \omega^2_p})-
 \frac{p(0){\dot{p}(0)}}{\kappa  \omega^2_p}
\label{C} \enq Integrating once more and fitting the constant so
as to satisfy the first initial condition in \er{ic}, we finally
obtain: \ber \Psi(t)& = & {1\over{2}}\ln\big[ \frac{
(1+\frac{{p^2}(t)}{\kappa \omega^2_p})}{ (1+\frac{{p^2}(0)}{\kappa
\omega^2_p})}\big] \nonumber
\\ &  & + \big[\frac{\Gamma_e}{2} (1+\frac{{p^2}(0)}{\kappa  \omega^2_p})-
\frac{p(0){\dot{p}(0)}}{\kappa   \omega^2_p}\big]
\int_0^t\frac{dt'} { 1+\frac{{p^2}(t')}{\kappa \omega^2_p}}
\label{Psif} \enr One can now substitute  this expression and
\er{Psidot}  into \er{edp2} to obtain, after considerable
simplification, an expression for the energy density arising from
the time varying "canonical" polarization  $p(t)$ in the form:
\beq (T^0_0)_{p,\Psi} = \frac{1}{2 \omega_p^2}\Bigl(\frac{{\dot
p}^2(t)+ \kappa \omega^2_p C^2}{1+\frac{p^2(t)}{\kappa
\omega^2_p}} +\omega_r^2 p^2(t)\Bigr) \label{ped3} \enq In the
case of slow relaxation, when  $\Gamma_e$ is small
(quantitatively, when $\Gamma_e \sqrt \kappa <<1$), then so is
$\dot p(0)$. (This will be confirmed in the next section.) This
results in  the quantity $C^2$  being a small quantity of the
second order,  which can be neglected: \beq (T^0_0)_{p,\Psi} =
\frac{1}{2 \omega_p^2}\Bigl(\frac{{\dot
p}^2(t)}{1+\frac{p^2(t)}{\kappa \omega^2_p}} +\omega_r^2
p^2(t)\Bigr) \label{ped3b} \enq For the part of the energy density
involving the "canonical" magnetization $m(t)$, the variables
$n(t)$ and $\Phi(t)$ having been eliminated through their
equations of motion, a similar procedure gives \beq
(T^0_0)_{m,\Phi} = \frac{1}{2F\omega_0^2}\Bigl(\frac{{\dot
m}^2(t)}{1+\frac{ m^2(t)}{\lambda F \omega^2_0}} +\omega_0^2
m^2(t)\Bigr) \label{med3} \enq The above two expressions for the
energy densities, \er {ped3b} and \er {med3}, resemble those in
equation (11) of  Ruppin \cite{Ruppin},
 except that they are written in the canonical (small letter) variables, rather than
 in the physical variables
(the relations between these will be presently obtained), and that
in the denominators  they contain the polarization fields squared.
The role of these is to damp out "kinetic" energy in the
polarization motion, associated with the motion of charged (or
spinning) matter.

The main results of this section, \er {ped3} and its analogue for
the magnetization energy density,  are exact and contain
non-perturbative corrections to the energy density, due to the
presence of the sink degrees of freedom. While exact, they are
model dependent in the sense that sinks represented by different
Lagrangians would lead to different energy  densities. This is
 clear, due to
the presence in the energy densities of the parameters $\kappa$ and $\lambda$
 that were introduced
 in the Lagrangian in \er{LagrPsi} and \er{LagrPhi}. This outcome
 was anticipated some time ago in \cite {Loudon}. (We also note the
 opposing view in
 \cite{Neufeld66}.)
It is of interest to note that the non-dissipative limit is not
regained when $\Gamma_e,\Gamma_h \to 0 $, but only when {\it also}
$\kappa, \lambda \to \infty$. It can be shown that in these limits
the canonical fields are identical to the physical fields.

\section{ Interpretation of the fields}

 We now find the relation of the canonical fields to the physical
 fields in the presence of dissipation and sink modes. We regain
 the original equations of motion, \er{eleom}
and \er{mageom}, for
 the physical polarization variable, as follows: We postulate
\ber
p(t) & = & e^{\Psi(t)} P(t)\\
m(t) & = & e^{\Phi(t)} M(t) \label{pPmM} \enr This turns (the
vector form of) \er{peom} into the following: \beq
 \ddot{\vec P} +2 {\dot \Psi}(t){\dot {\vec P}}(t) + \omega^2_r {\vec P} =
  \omega^2_p e^{-\Psi(t)}{\vec e}
  \label {neweleom2}
  \enq
and likewise for the magnetization variables. Then, from \er{ic},
for short times $0<t<< \frac{2}{\Gamma_e}$, \beq {\dot\Psi}(t)
\approx
 \dot \Psi(0)  =\Gamma_e /2 \label{initPsi}
\enq We then obtain \beq \ddot{\vec P} +\Gamma_e {\dot {\vec
P}}(t) + \omega^2_r {\vec P} =
   \omega^2_p e^{-\Psi(t)}{\vec e}
  \label {neweleom}
  \enq  Recalling \er{eleom} and
 \er {mageom},  we can thus extrapolate to later times so as to identify
\ber
{\vec e(t)} & = & e^{\Psi(t)} {\vec E(t)}\\
{\vec h(t)} & = & e^{\Phi(t)} {\vec H(t)} \label{eEmM} \enr and
regain \er{eleom} and \er{mageom}. This provides a physical
meaning for {\it all} the "canonical" variables, as those fields
in which the decay has been reinstated. On the other hand, the
  decay is itself dependent
 on the fields. (Cf. \cite {Neufeld69}.) Furthermore, the Maxwell equations for the physical
 fields are also modified in the dissipative-polarizable medium. This feature (of a {\it modified} Maxwell equation) also appears
 in  {\cite {Bekenstein2002} and   \cite {Bekenstein1982} (eq. (14) and  eq. (12), respectively).

\section {A covariant dissipative Lagrangian density}

In this section we introduce a covariant Lagrangian density which
is a Lorentz invariant generalization of the Lagrangian density
described in \er{Lagrangian}. Again we introduce a set of new
fields, which are formally distinguished from
 the previous, physical fields by writing them in lower case  symbols.
 Explicitly, we shall work with  the scaled polarization and electromagnetic tensors
 $p^{\alpha \beta}$ and $f^{\alpha \beta}$ and other lower case quantities that were introduced
 earlier. Notice
that the Lagrangian introduced here will contain less terms since both the magnetization
and the polarization are now part of the polarization tensor $p^{\alpha \beta}$.
This Lagrangian has the following parts:
  \beq
  {\cal L} = {\cal L}_{fp} + {\cal L}_{p\Psi} + {\cal L}_{\Psi}
  \label {Lagrangian2}
  \enq
In which ${\cal L}_{fp}$ is defined in \er{staticlag2} in the appendix (although
an upper case  symbols to lower case  symbols transformation is
needed). ${\cal L}_{p\Psi}$ is obtained by generalizing
\er{lagrfirst} as follows: \beq {\cal L}_{p\Psi} = \frac{1}{2
\hat{\omega_p^2}} \left[\partial_\mu p^{\alpha \beta}
\partial^\mu p_{\alpha \beta} - (\hat{\omega_r^2}- \partial_\mu \Psi \partial^\mu \Psi )
p_{\alpha \beta} p^{\alpha \beta} - 2 p_{\alpha \beta}
\partial_\mu p^{\alpha \beta} \partial^\mu \Psi \right]
\label{LPPsi} \enq ${\cal L}_{p\Psi}$ can be written as a sum of
two terms: one depending on the magnetization, while the other
depends on the polarization. \ber {\cal L}_{p\Psi} &=& {\cal
L}_{\vec p\Psi} + {\cal L}_{\vec m \Psi}
\nonumber \\
{\cal L}_{\vec p\Psi} & = & \frac{1}{\hat{\omega_p^2}} [ -\frac{1}{c^2}(\partial_t \vec p)^2 + (\partial_i \vec p)^2
+(\hat{\omega_r^2}- \frac{1}{c^2} (\partial_t \Psi)^2 + (\vec \nabla \Psi)^2 )\vec p^2
\nonumber \\
& + &
\frac{2}{c^2} \vec p \cdot \dot{\vec p} \dot \Psi - 2 \vec p \cdot \partial_i \vec p \partial_i \Psi ]
\nonumber \\
{\cal L}_{\vec m\Psi} & = & - \frac{1}{\hat{\omega_p^2}} [ -\frac{1}{c^2}(\partial_t \vec m)^2 + (\partial_i \vec m)^2
+(\hat{\omega_r^2}- \frac{1}{c^2} (\partial_t \Psi)^2 + (\vec \nabla \Psi)^2 )\vec m^2
\nonumber \\
& + & \frac{2}{c^2} \vec m \cdot \dot{\vec m} \dot \Psi - 2 \vec m
\cdot \partial_i \vec m \partial_i \Psi ] \enr Finally ${\cal
L}_{\Psi}$ is obtained by generalizing \er{LagrPsi} as follows:
\beq {\cal L}_{\Psi} = \frac{1}{2} [\hat \kappa \partial_\mu \Psi
\partial^\mu \Psi -  \rho \Psi^2] = \frac{1}{2} [\hat \kappa
(\frac{1}{c^2} (\partial_t \Psi)^2 - (\vec \nabla \Psi)^2) -  \rho
\Psi^2] \enq The field \eqs with respect to $p^{\alpha
\beta},\Psi$ and the electromagnetic fields can be obtained by
taking the variational derivative of  ${\cal L}$ given in
\er{Lagrangian2} with respect to $p^{\alpha \beta},\Psi$ and
$a^\alpha$. The electromagnetic field \eqs were already derived by
taking the variational derivative of \er{staticlag2} with respect
to $a^\alpha$ and further change is unnecessary (except for that
 from upper to lower case symbols ). Taking the variational
derivative of ${\cal L}$ with respect to $\Psi$ leads to the
equation: \beq \hat \kappa \partial_\mu
\partial^\mu \Psi + \rho \Psi =\frac{1}{\hat{\omega_p^2}}
\partial^\mu [p^{\alpha \beta}(\partial_\mu p_{\alpha \beta} - \partial_\mu \Psi p_{\alpha \beta})]
\label{Psie}
\enq
Taking the variational derivative of ${\cal L}$ with respect to $p_{\alpha \beta}$ leads to the equation:
\beq
\partial_\mu \partial^\mu p_{\alpha \beta} + (\hat{\omega_r^2}- \partial_\mu \Psi \partial^\mu \Psi
- \partial_\mu \partial^\mu \Psi ) p_{\alpha \beta}  = -
\frac{1}{2} \hat{\omega_p^2} f_{\alpha \beta} \label{Palbe} \enq
In the "homogeneous" case in which the spatial derivatives of both
the polarization tensor $p^{\alpha \beta}$ and the scalar field
$\Psi$ vanish, the tensor \er{Palbe} can be written as two vector
equations:
 \ber
    \ddot{\vec p}-( \ddot{\Psi}+{\dot{\Psi}}^2 -\omega^2_r)\vec p & = & \omega^2_p \vec e
    \label{peom2}\\
    \ddot{\vec m}-( \ddot{\Psi}+{\dot{\Psi}}^2 -\omega^2_r)\vec m & = & - \omega^2_p \vec b
    \label{meom2}
 \enr
In which $\omega^2_r = c^2 \hat \omega^2_r$ and $\omega^2_p =
\frac{1}{2} c^2 \hat \omega^2_p$. We can clearly see that
\er{peom2} is the same as \er{peom}. As for \er{meom2} by using
the equality (\ref{HB}) this can be rewritten as: \beq \ddot{\vec
m}-( \ddot{\Psi}+{\dot{\Psi}}^2 -\omega^2_0)\vec m = - \omega^2_p
\vec h \enq In which $\omega^2_0 = \omega^2_r + 4 \pi \omega^2_p
$. This \eq \ can be identified with \er{meom} by choosing $F=
-\frac{\omega^2_p}{\omega^2_0}$ and equating the magnetic and
electric dissipative modes ($ \Phi\equiv\Psi$ ). Notice that in
magnetized materials the magnetic field is in a direction opposite
to the magnetization as can be seen from figure \ref{mf}, which
explains the negative sign in the above equation.
\begin{figure}
\vspace{3.5cm} \includegraphics{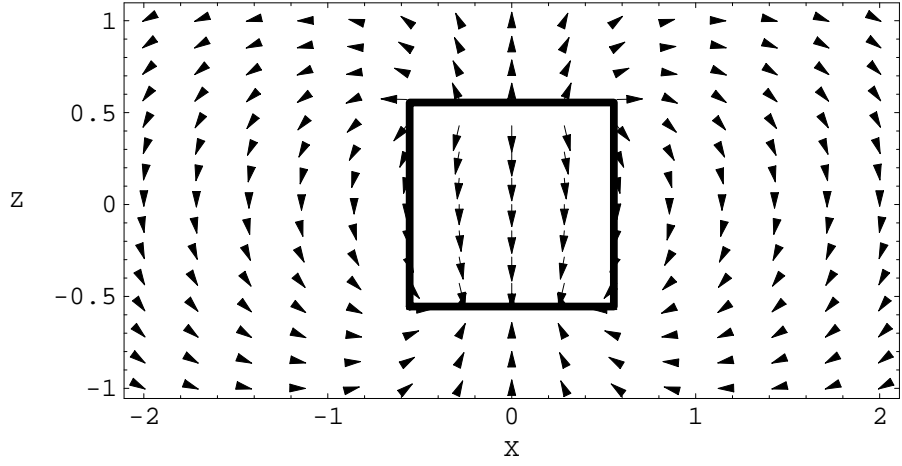} \caption {Magnetic field lines in a box of
permanent uniform magnetization - a cross section view. The direction of
magnetization can be inferred from the way the magnetic field
lines look {\bf outside} the box magnet. The normal component of
the field changes sign across the material boundary, unlike in the
dielectric case.} \label {mf}
\end{figure}

\subsection {The energy momentum tensor of the dissipative \\ Lagrangian density}

Using the dissipative Lagrangian density defined in
\er{Lagrangian2} we can calculate the energy momentum tensor by
using the formula given in (\ref{Tkj}). Doing so we obtain the
following expression: \beq T_\alpha^\beta = T_\alpha^\beta [fp] +
T_\alpha^\beta [p\Psi] + T_\alpha^\beta [\Psi] \label{enmotdis}
\enq $T_\alpha^\beta [fp]$ is calculated in Appendix (A.3) in
\er{enmotefp}. It remains to obtain expressions for
$T_\alpha^\beta [p\Psi]$ and $T_\alpha^\beta [\Psi]$; these are
given below: \ber T_\alpha^\beta [p\Psi]&=&\frac{1}{\hat
\omega^2_p} [
\partial^\beta p^{\gamma \theta} \partial_\alpha p_{\gamma \theta}
- p^{\gamma \theta} (\partial^\beta \Psi
\partial_\alpha p_{\gamma \theta} +
\partial_\alpha \Psi \partial^\beta p_{\gamma \theta} )
\nonumber \\
&+&  \partial^\beta \Psi \partial_\alpha \Psi p_{\gamma \theta}
p^{\gamma \theta}] - \delta^\beta_\alpha {\cal L}_{p\Psi}
\label{enmotPPsi} \enr and \beq T_\alpha^\beta [\Psi]= \hat \kappa
\partial^\beta \Psi \partial_\alpha \Psi - \delta^\beta_\alpha
{\cal L}_{\Psi} \label{enmotPsi} \enq Let us look at the energy
density $T_0^0$. For $T_0^0 [\Psi]$ we obtain the positive
quantity \beq T_0^0 [\Psi]= \frac{1}{2}[\frac{\hat \kappa}{c^2}
\dot\Psi^2 + \hat \kappa (\vec \nabla \Psi)^2 + \rho \Psi^2]
\label{enmot00Psi} \enq For $T_0^0 [p \Psi]$ we obtain \beq T_0^0
[p\Psi]= \frac{1}{\hat \omega^2_p} [(\partial_\mu \vec m -
\partial_\mu \Psi \vec m)^2- (\partial_\mu \vec p - \partial_\mu
\Psi \vec p)^2 + \hat \omega^2_r (\vec m^2 - \vec p^2)]
\label{enmot00PPsi} \enq This expression contains positive
magnetization contributions and negative polarization
contributions. The positive magnetization contributions can
overcome the negative magnetization contribution of $T_0^0 [fp]$
given in \er{enmote00fp} of Appendix (A.3), provided that
$\frac{\hat \omega^2_r}{\hat \omega^2_p} > 2 \pi $. However, the
negative polarization contribution to $T_0^0$ results in the
uncertainty whether the total polarisation contribution is
negative or positive, since this depends on the solution of \eqs
(\ref{Maxinhom4D},\ref{Psie},\ref{Palbe}). Those \eqs are
difficult to solve since they are complicated non-linear partial
differential equations. In the following subsection we will
present a solution for a particular, simple situation. Notice that
this situation is completely different from the situation in the
non-covariant model for which the positiveness of the energy
density was assured. To conclude this subsection we calculate the
contributions to $T_0^i$: \beq T_0^i = T_0^i [fp] + T_0^i [p\Psi]
+ T_0^i [\Psi] \enq since $T_0^i [fp]$ given in \er{Poynting} is
the Poynting vector, we can consider $T_0^i [p\Psi]$ and $ T_0^i
[\Psi]$ as corrections to the Poynting vector. Those are given by:
\ber T_0^i [p\Psi] &=& \frac{2}{c \hat \omega_p^2}[\vec \nabla p^k
\dot p^k - \vec \nabla m^k \dot m^k - p^k (\vec \nabla \Psi \dot
p^k + \dot \Psi \vec \nabla p^k)
\nonumber \\
&+& m^k (\vec \nabla \Psi \dot m^k + \dot \Psi \vec \nabla m^k) -
\vec \nabla \Psi \dot \Psi (\vec m^2 - \vec p^2)]
\label{enmote0iPPsi} \enr and \beq T_0^i [\Psi] = - \frac{\hat
\kappa}{c} \vec \nabla \Psi \dot \Psi \label{enmote0iPsi} \enq One
should notice that both these corrections vanish for the
homogeneous case in which the spatial derivatives vanish.

\subsection {A covariant solvable case}

In this final subsection we would like to elucidate the nature of
the energy density component $T^0_0$ by introducing a simple
solution of \eq \ (\ref{Psie}). We assume a homogeneous situation
in which all spatial derivative vanish. Furthermore, we assume the
$\rho =0$ and introduce the constant $\kappa = \frac{\hat
\kappa}{c^2}$. In this case we obtain the equation \beq \kappa
\ddot \Psi = \frac{1}{\omega_p^2} \partial_t[\vec m \cdot \dot
{\vec m} - \dot \Psi \vec m^2 - \vec p \cdot \dot {\vec p} + \dot
\Psi \vec p^2  ] \label{psieqsim} \enq From this equation we can
obtain an expression for $\dot \Psi$, as follows: \beq \dot \Psi =
(\frac{1}{2}) \frac{\partial_t (\vec m^2 - \vec p^2)}{\kappa
\omega_p^2 + \vec m^2 - \vec p^2} \label{dotpsisim} \enq (In which
we have neglected a constant that can be considered small). Using
\er{dotpsisim} we can rewrite \er{enmot00Psi} as \beq T_0^0
[\Psi]= \frac{1}{2} \kappa  \dot\Psi^2 = \frac{1}{8} \kappa
\left(\frac{\partial_t (\vec m^2 - \vec p^2)}{\kappa \omega_p^2 +
\vec m^2 - \vec p^2}\right)^2 \label{homenmot00Psi} \enq To this
we need to add the homogeneous energy density \beq T_0^0 [p\Psi]=
\frac{1}{2 \omega_p^2} [(\dot{\vec m} - \dot \Psi \vec m)^2-
(\dot{\vec p} - \dot \Psi \vec p)^2 + \omega^2_r (\vec m^2 - \vec
p^2)] \label{homenmot00PPsi} \enq We will consider to separate
cases:
\begin{enumerate}
\item $\vec p =0, \vec m = m \hat x$. In this case we obtain \beq
T_0^0 [\Psi] + T_0^0 [p\Psi]= \frac{1}{2 \omega_p^2} [ \omega^2_r
m^2 + \frac{\dot m^2}{1+\frac{m^2}{\kappa \omega_p^2}}] \enq which
is very similar to equation (\ref{med3}). This expression is
positive and with the correct choice of parameters can overcome
the negative magnetization contribution of $T_0^0 [fp]$ \item
$\vec p =p \hat x, \vec m = 0 $. In this case we obtain \beq T_0^0
[\Psi] + T_0^0 [p\Psi]= \frac{1}{2} [\frac{\kappa \dot p^2 }{p^2 -
\kappa \omega_p^2} - \frac{\omega_r^2}{\omega_p^2} p^2] \enq which
is very different from equation (\ref{ped3b}). This expression is
positive only if $p^2 > \kappa \omega_p^2$ and if $p$ changes fast
enough to overcome the negative part
$-\frac{\omega_r^2}{\omega_p^2} p^2$; otherwise, the energy
density is negative.
\end{enumerate}

 \section{Conclusion}

In this paper the Lagrangian method for a dissipative medium
 (capable also  of sustaining  electric  and/or magnetic polarizations)
 was carried out to obtain equations of motion, energy-momentum
 densities, etc. To apply the method
 it has been found necessary to introduce additional degrees of freedom
 ("sink-modes"), associated with  decay mechanisms in the electrical and magnetic
 modes.

  Two versions have been formulated: The first one, which is not invariant under
 Lorentz transformation, has led to equations
for both the electromagnetic fields (a slightly corrected set of Maxwell equations) , polarization and
magnetization and, using a model for the sink variables, has
unambiguously yielded energy densities, shown in \er{ped3} -
\er{med3}. These are  positive and resemble the corresponding
results in \cite {Ruppin}.
 The results obtained are similar to the ones obtained in \cite{EnglmanY}, however,
 they have been obtained here with a much simpler Lagrangian, containing fewer degrees of freedom.
 To be precise, two vector degrees of freedom (or six independent quantities)  have been
 omitted from the beginning, thus making the formalism more  useful for numerical calculations.

Secondly, a  covariant Lagrangian was formulated, one that is
invariant under the Lorentz transformation and contains spatial as
well as temporal derivatives. From this, we have derived equations
for both the electromagnetic fields, polarization and
magnetization . Calculating the energy momentum tensor we have
derived expression for both the energy density and the generalized
Poynting vector. However, we find that with a specific model for
the sink modes, the energy density derived from this covariant
Lagrangian is not positive definite. This points to the need to
introduce a better model for the sink degrees of freedom, perhaps
with better physical underpinning.
\\
\\
{\Large \bf APPENDIX}

\appendix

\section{The covariant formalism}

  In a covariant formalism of the electromagnetic theory one uses a four dimensional
  space-time formalism instead of the usual distinction that is practised between spatial
  and temporal coordinates. In this formalism the invariance of the Lagrangian of the
  electromagnetic theory under Lorentz transformation is clear.

  \subsection{Four dimensional notations}

  The four dimensional coordinate are defined as:
  \beq
    x^0 = c t, \quad x^1=x, \quad  x^2=y,  \quad x^3 =z
  \label{4Dcor}
  \enq
  In this formalism there is a difference between coordinates with upper and lower indexes.
   For example:
   \beq
    x_0 = c t, \quad x_1=-x, \quad  x_2=-y,  \quad x_3 = - z
  \label{4Dcorb}
  \enq
  The connection between upper and lower coordinates is established through the metric
   matrix $g_{\alpha \beta}$, which is defined as
  \beq
g_{\alpha \beta} = g^{\alpha \beta} = {\bf diag} (1,-1,-1,-1)
  \enq
 (we shall use Greek letters for 4 coordinates according to the well established custom) such that
 \beq
g_{\alpha \gamma}  g^{\gamma\beta} = \delta_{\alpha}^{\beta}={\bf diag} (1,1,1,1)
  \enq
Using $g_{\alpha \beta}$ we can write $x_{\alpha}=g_{\alpha \beta}
x^{\beta}$. This notation will be adapted for any four vector \beq
A^\alpha = (A^0, \vec A), \quad A_\alpha = (A^0, -\vec A), \quad
A_{\alpha}=g_{\alpha \beta} A^{\beta} \enq For the four
dimensional derivatives we will use the notation \beq
\partial^\alpha = \frac{\partial}{\partial x_{\alpha}} = (\frac{\partial}{\partial x^0},- \vec \nabla),
\quad \partial_\alpha = \frac{\partial}{\partial x^{\alpha}} = (\frac{\partial}{\partial x^0}, \vec \nabla)
\enq

\subsection{Electromagnetic theory in four dimensions}

The vector and scalar electromagnetic potentials are replaced by a
single four-vector \beq A^\alpha = (\Phi, \vec A) \label{4Dpot}
\enq Using \eqs (\ref{pot}) and (\ref{4Dpot}) we can define the
contravariant and covariant antisymmetirc tensors: \beq
F^{\alpha \beta} = \partial^\alpha A^\beta - \partial^\beta A^\alpha=\left(%
\begin{array}{cccc}
  0 & -E^x & -E^y & -E^z \\
  E^x & 0 & -B^z & B^y \\
  E^y & B^z & 0 & -B^x \\
  E^z & -B^y & B^x & 0 \\
\end{array}%
\right)
\label{Fup}
\enq
\beq
F_{\alpha \beta} = g_{\alpha \gamma}F^{\gamma \eta} g_{\eta \beta}=
\left(%
\begin{array}{cccc}
  0 & E^x & E^y & E^z \\
  -E^x & 0 & -B^z & B^y \\
  -E^y & B^z & 0 & -B^x \\
  -E^z & -B^y & B^x & 0 \\
\end{array}%
\right) \label{Fdown} \enq Analog tensors can also be defined for
the magnetic field $\vec H $ and electric displacement field $\vec
D$ such that \beq
G^{\alpha \beta} = \left(%
\begin{array}{cccc}
  0 & -D^x & -D^y & -D^z \\
  D^x & 0 & -H^z & H^y \\
  D^y & H^z & 0 & -H^x \\
  D^z & -H^y & H^x & 0 \\
\end{array}%
\right),
G_{\alpha \beta} = \left(%
\begin{array}{cccc}
  0 & D^x & D^y & D^z \\
  -D^x & 0 & -H^z & H^y \\
  -D^y & H^z & 0 & -H^x \\
  -D^z & -H^y & H^x & 0 \\
\end{array}%
\right) \label{G} \enq Finally we introduce the polarization
tensors which contain both the magnetization $\vec M$ and
polarization $\vec P$ \beq
P^{\alpha \beta} = \left(%
\begin{array}{cccc}
  0 & P^x & P^y & P^z \\
  -P^x & 0 & -M^z & M^y \\
  -P^y & M^z & 0 & -M^x \\
  -P^z & -M^y & M^x & 0 \\
\end{array}%
\right),
P_{\alpha \beta} = \left(%
\begin{array}{cccc}
  0 & -P^x & -P^y & -P^z \\
  P^x & 0 & -M^z & M^y \\
  P^y & M^z & 0 & -M^x \\
  P^z & -M^y & M^x & 0 \\
\end{array}%
\right) \label{P} \enq In terms of \eqs (\ref{Fup}), (\ref{G}) and
(\ref{P}) we can rewrite \er{DE} and \er{HB} as a single tensor
equation \beq G^{\alpha \beta} = F^{\alpha \beta} - 4 \pi
P^{\alpha \beta} \enq Furthermore, the \eqs appearing in
(\ref{Maxinhom}) can be rewritten as a single tensor equation \beq
\partial_\alpha G^{\alpha \beta} = \frac{4 \pi}{c} J_f^\beta
\label{Maxinhom4D} \enq In this the four current is defined as
$J^\alpha = (c \rho, \vec J)$ and the subscript {\it f} \ refers
to free currents and charges. In what follows we assume that the
material under study does not contain any {\bf free} currents or
charges. The covariant Lagrangian density for the electromagnetic
field can be written as \ber {\cal L}_{FP} &=& {\cal L}_{F} +
{\cal L}_{FJ} \nonumber
\\
{\cal L}_{F} &=& -\frac{1}{16 \pi} F_{\alpha \beta} F^{\alpha \beta}
\nonumber
\\
{\cal L}_{FJ} &=& -\frac{1}{c} J_{\alpha} A^{\alpha}
\label{staticlag2} \enr In matter $J_\alpha$ can be written as
\beq J_{\alpha} = c \partial^\beta P_{\beta \alpha} \enq This
expression automatically satisfies the conservation law
$\partial^\alpha J_\alpha =0$, due to the antisymmetry of
$P_{\beta \alpha}$. The field \eqs (\ref{Maxinhom4D}) can be
derived by taking the variational derivative of \er{staticlag2}
with respect to the four potential $A^\alpha$. Writing ${\cal L}$
given in \er{staticlag2} in terms of the standard notation leads
to the expression \beq {\cal L}_{FP} = \frac{1}{8 \pi}[\vec E^2 -
\vec B^2]+ \vec P \cdot \vec E + \vec M \cdot \vec B + \vec \nabla
\cdot (\Phi \vec P + \vec M \times \vec A) + \frac{1}{c}\partial_t
(\vec A \cdot \vec P) \label{staticlag3} \enq Notice the
differences between \er{staticlag} and the covariant result given
in \er{staticlag3}:
\begin{enumerate}
    \item The covariant formalism contains the divergence term $\vec \nabla \cdot (\Phi \vec P + \vec M \times \vec A)$
    which is absent in the previous formalism. This term will contribute only on the boundary of the
    domain of integration and  can be ignored if we assume that the domain of integration is over the  entire
    space and the fields vanish at infinity.
    \item The covariant formalism contains the total time derivative
    $\frac{1}{c}\partial_t (\vec A \cdot \vec P)$ which is absent in the previous formalism.
    This term will not contribute to the field equations and can be ignored.
    \item The coupling between the magnetization and magnetic field is different. In the covariant formalism
    the induction density $\vec B$ is coupled to the magnetization instead of the magnetic field $\vec H$.
\end{enumerate}

\subsection{The energy momentum tensor of the electromagnetic field}

The energy momentum tensor in terms of arbitrary fields
$\eta_\rho$ is given by \beq T_\alpha^\beta = \frac{\partial {\cal
L}}{\partial (\partial_\beta \eta_\rho)}\partial_\alpha \eta_\rho
- \delta _\alpha^\beta \cal L \label{enmote} \enq For the
electromagnetic field, when we  calculate the energy momentum
tensor $T_\alpha^\beta [FP]$ using ${\cal L}_{FP}$, this results
in \beq T_\alpha^\beta [FP] = \frac{1}{4 \pi}
\partial_\alpha A_\mu F^{\mu \beta} + A_\mu \partial_\alpha P^{\mu
\beta} - \delta _\alpha^\beta {\cal L}_{FP} \label{enmotefp} \enq
The energy density is given by $T_0^0 [FP]$ which can be written
as follows: \beq T_0^0 [FP] = \frac{1}{8 \pi} [\vec E^2 + \vec H^2
- (4 \pi \vec M)^2]+ \vec \nabla \cdot (\frac{\Phi \vec E}{4 \pi}
+ \vec A \times \vec M) \label{enmote00fp} \enq The divergence
term $\vec \nabla \cdot (\frac{\Phi \vec E}{4 \pi} + \vec A \times
\vec M)$ will only contribute as a boundary term to the
electromagnetic energy and can be ignored if we assume that the
domain of integration is over the entire space and the fields
vanish at infinity. Notice that for the covariant Lagrangian
density the energy density is different from the result of
\er{edstatic}: it contains a negative quadratic term in the
magnetization instead of a positive one. If the magnetization is a
given static quantity, this is just a constant that can be
ignored. On the other hand, if the magnetization is a dynamic
degree of freedom, this term can render the energy negative. Of
course, if the magnetization is dynamic, it will have its own
Lagrangian which will have its own contribution to the energy
momentum tensor; this will be discussed in the following
subsections.

 (One would expect that the relativistic energy density in \er{enmotefp}
 would smoothly approach the non-relativistic expression in
 \er{edstatic} as $c\to \infty$. Apparently this is not the case,
 because of the different signs of the quadratic magnetization
 term. In truth, when one goes deeper, into the atomic theory of
 magnetization, one recognizes that both the orbital and spin
 contributions to the magnetization $M$ are proportional to the
 Bohr magneton $\frac{e\hbar}{2mc}$. This vanishes in a theory
 where $c\to \infty$.)

 Finally we derive the Poynting vector
 by calculating $T_0^i [FP]$. This results in
\beq
T_0^i [FP] =[ \frac{1}{4 \pi} \vec E \times \vec H +
\frac{1}{c}\partial_t (\vec M \times \vec A - \frac{\Phi \vec E}{4 \pi} )+
\vec \nabla \times (\frac{\Phi \vec H}{4 \pi})]^i
\label{Poynting}
\enq
which is the "correct" form if we ignore boundary terms and total
time derivatives.

\end{document}